# Soft mode parameter as an indicator for the activation energy spectra in metallic glass

Shan Zhang [a], Chaoyi Liu[b], Yong Yang[c], Yue Fan [b], Pengfei Guan [a,*]

[a] Beijing Computational Science Research Center, Beijing, China

[b] Department of Mechanical Engineering, University of Michigan, Ann Arbor, USA

[c] Centre for Advanced Structural Materials, Department of Mechanical and Biomedical Engineering, City University of Hong Kong, Hong Kong, China

Corresponding author: Pengfei Guan[*]

Email: pguan@csrc.ac.cn

**ABSTRACT:** The activation energy ($E_A$) spectra of potential energy landscape (PEL) provides a convenient perspective for interpreting complex phenomena in amorphous materials; however, the link between the $E_A$ spectra and other physical properties in metallic glasses is still mysterious. By systematically probing the $E_A$ spectra for numerous metallic glass samples with distinct local geometric ordering, which correspond to broad processing histories, it is found that the shear modulus of the samples are strongly correlated with the arithmetic mean of the $E_A$ spectra rather than with the local geometrical ordering. Furthermore, we studied the correlation of the obtained $E_A$ spectra and various well-established physical parameters. The outcome of our research clearly demonstrates that the soft mode parameter $\Psi$ and the $E_A$ spectrum are correlated; therefore, it could be a good indicator of metallic glass properties and sheds important light on the structure-property relationship in metallic glass through the medium of PEL.

**TOC Graphic**

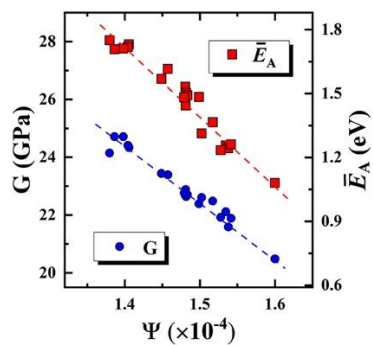

Metallic glasses (MGs), as metastable systems, can provide ample opportunity for property control that can be used in structural and functional applications [1-8]. However, due to their inherent nature of disordered atomic packing and, consequently, the absence of conventional defects with distinctive topologies (*e.g.* dislocations, grain boundaries, *etc.*), it is very difficult to establish a "crystal clear" structure-property relationship for MGs. In fact, the definition of "defects" in disordered materials are obscure and not unique: examples include free volume (FV) [9-10], shear transformation zones (STZ) [11-15], Voronoi tessellation [16-17], flow units [18-20], soft modes/spots [21-23], regions with large fluctuations of atomic level stresses [24-26], *etc*. Tremendous efforts have been made in the past to examine the correlations between the local atomic relaxations and those "defects" [21, 27-30]. On the other hand, it is known that many macroscopic properties of MGs (*e.g.* heat release/absorption [31-32], homogeneous deformation [18], *etc.*) are actually governed by the overall structural distributions and activation energy ($E_A$) spectra, rather than by some extreme "defects", which are located near the ends of the spectra. From the PEL perspective, the $\alpha$ relaxation process can be pictured as the transitions between the deep 'metabasins', whereas the $\beta$ relaxation process refers to the elementary hopping event between the 'sub-basins' coded within a metabasin. These processes are related to many important properties, such as glass transition, deformation, aging and rejuvenation etc., of MGs. Systematically analyzing the activation energy ($E_A$) spectra for these processes are extremely helpful for understanding some critical phenomena[33], and establishing the microscopic structure-property connections. Therefore, important questions that follow and yet remain open are: how does the $E_A$ spectra correlate with the physical properties of interest in MGs, and what are the most effective structural indicators in determining MGs' $E_A$ spectra?

In this work, molecular dynamic (MD) simulations that combine potential energy landscape

(PEL)-sampling technique were employed to investigate the physical properties and systematically probe the $E_A$ spectra for numerous MGs samples with distinct local geometry or short-range order (SRO) behaviors. These samples are prepared over broad processing histories in the cooling rate—pressure space. It is found that the shear modulus G is strongly correlated with the arithmetic mean of $E_A$ spectra rather than with the local geometry ordering (such as the fraction of icosahedron clusters), even though the processing history does induce pronounced local structure changes. Focuses are placed on how those $E_A$ spectra are correlated with the five widely used representative parameters, including the overall density, the inherent structure energy ($E_{IS}$), the two-body entropy ($S_2$), the soft-mode parameter ($\Psi$), and the local configurational anisotropy ($|\vec{u}|$). It is found that a sample's $E_{IS}$ and soft-mode parameter ($\Psi$) exhibit the best correlation with the arithmetic mean $\bar{E}_A$. By further quantifying the differences between each pair of obtained $E_A$ spectra and mapping them into various measure spaces, we demonstrate that, among those five representative parameters, the soft-mode parameter performs the best in describing a sample's $E_A$ spectrum and, thereby, the shear modulus G.

A $Cu_{50}Zr_{50}$ model system with the realistic embedded atom method (EAM) potential [17, 34] is considered in the present work as a prototypical MG. The details of our model systems and activation-relaxation technique (ART) method are presented in the Simulation Methods and Supporting Information. In brief, the 4000-atom glass samples are prepared by systematically tuning the hydrostatic stress (0, 5, 10, 15, 20 GPa) and the cooling rate ($10^{10}$, $10^{11}$, $10^{12}$, $10^{13}$ K/s). To be more specific, the preparation process for each sample consists of two steps (**Fig 1a**): (i) quench the initial supercooled liquid at the prescribed pressure ($P_i$) and cooling rate condition to 1K; (ii) release the applied pressure and let the system relax at 1K for 10ns. The state of metallic glasses is tuned by the cooling rate and prescribed pressure $P_i$ during the quenching process, even the density changes are slight (~1%). Both the cooling rate and the $P_i$ introduce changes to the total radial distribution functions (**Fig. S1**), but the influence of $P_i$ is more pronounced than cooling rate. This indicates that the atomic structure, especially the local atomic packing behavior of these samples should be different. As shown in **Fig. 1b**, the fractions of <0,0,12,0> cluster (denoted as $f_{<0,0,12,0>}$) of samples increases nearly three times due to the high-pressure effect, which is comparable to the results in ref. [35]. The $f_{ico}$ is tuned in a wide range (0.02~0.12) by varying $P_i$ and the cooling rate. This behavior suggests that these MG samples have distinct local geometries or short-range orders (SROs). Therefore, these samples form a system with similar densities but vastly different atomic configurations. Their physical properties, such as the shear modulus G and the activation energy ($E_A$) spectra etc. will be investigated to reveal the correlations, if any, between the structural features and the physical properties of interest.

Since the shear modulus is at the heart of may key properties of MGs, we focus on the correlation between G and other parameters. Figure 1c shows the correlation between the $f_{<0,0,12,0>}$ and the calculated shear modulus G of the investigated samples. It is clear that we do not obtain the collapse of data for these samples. This means that the local geometrical order is not appropriate for describing the mechanical properties of MGs, as mentioned in the previous studies[35-36]. According to the PEL perspective, we may obtain the collapse of data G as a function of the intrinsic behaviors of the $E_A$ spectra, such as the arithmetic mean of the $E_A$ spectra, $\bar{E}_A$. By performing the ART in our MD simulations, the $E_A$ spectra of 20 samples in Fig. 1b can be monitored. The related arithmetic mean $\bar{E}_A$ is calculated by $\bar{E}_A = \langle \sum_{i=1}^{N_{event}} E_{A,i} \rangle$, where $N_{event}$ is the number of valid events in the ART methord(see the Supporting Information for more details ). As we plotted G as a function of $\bar{E}_A$ (Fig. 1d), it is amazing to find the collapse of data for all samples. In theory, this behavior provides the direct evidence for the intrinsic correlation between the PEL and mechanical properties. In concept, this finding also implies that the systematical study of PEL may build a bridge between the structure and properties of MGs on a quantitative basis. Now, the question is: Is there a structural parameter that is correlated with the $E_A$ spectra.

To address the above question, we analyzed the $E_A$ spectra of each sample. **Fig. 2a** shows the $E_A$ spectra of 20 samples considered in the present study, and the corresponding cooling rates and hydrostatic stresses are also marked in the plot. It can be seen that, as the cooling rate increases (*i.e.* comparing the panels in the same row), the $E_A$ distribution gradually shifts towards the left side and develops a fat tail in the low $E_A$ region. This is not surprising, because a higher cooling rate means the sample has less time to relax and therefore is quenched in a less stable state that is more susceptible to perturbations [32, 37]. On the contrary, the pressure effects (*i.e.* comparing the panels in the same column) are much more complex and seem to couple with the cooling rates. For example, under the cooling rate of $10^{10}$K/s, the $E_A$ spectra are insensitive to the pressure; while under the cooling rate of $10^{13}$K/s, the $E_A$ spectra clearly shift towards the low-$E_A$ end as the pressure increases. This is somewhat surprising because, intuitively, a higher pressure would squeeze out free volumes, thus making the sample more stable. It is worth noting that such abnormal dependence on pressure is consistent with the recent studies [38], which is an interesting topic for future investigation.

To unravel the origin of the complex $E_A$ spectra, we construct the radar plots, as seen in the inset of each panel, which show the relative importance of 5 widely used structural indicators, including the density ($\rho$), the inherent structure energy ($E_{IS}$), the two-body entropy ($s_2$), the soft-mode parameter ($\Psi$), and the local configurational anisotropy ($|\vec{u}|$). We would like to note that, while these structural indicators can be defined locally at atomic level (see the Supporting Information), here we focus on their average values in the radar plots and also in the following discussions in line with the global nature of the $E_A$ spectrum.

We first examine how the above-mentioned 5 indicators of a sample correlate with the simple arithmetic mean $\bar{E}_A$. To better compare the different measures, a polar coordinate system is employed, and all the 5 panels are presented in the same plot in **Fig. 2b**. The radial axis represents the magnitude of the arithmetic mean activation energy, $\bar{E}_A$; while the [0º,180º] angular regime is equally divided into five 36º-wide sub-regimes to represent different structural indicators. To be more specific, the [144º,180º] regime represent ρ, which is usually used to estimate the free volume of MGs in experiments [39]. As discussed above, it is intuitive that a higher density should correspond to less free volume and, consequently, a larger effective $\bar{E}_A$. However, no such correlation can be observed in the present study and the data points are widely scattered. In the [108º,144º] regime we consider $s_2$, which was shown in some prior studies as an effective indicator for slow kinetics [40-41]. Although the $\bar{E}_A$ seems to become larger as $s_2$ decreases, which is consistent with the previous studies [40], yet the overall correlation is rather weak. In the [72º,108º] regime, we consider $|\vec{u}|$, which has drawn much attention in recent studies [42-44] due to its predictive power for particle rearrangements. As seen by the blue hexagons in **Fig. 2b**, indeed there is a tendency that a larger $|\vec{u}|$ would increase the likelihood of atom rearrangements (*i.e.* lower $\bar{E}_A$.) But still, the data show a significant scatteredness, suggesting that the correlation is not strong. In the [36º,72º] regime we examine $E_{IS}$, which has also been widely used as an indicator of glasses' stability [32, 37, 45-46]. It can be seen that the data points are reasonably well aligned with a diagonal curve, indicating a linear dependence between the two variables in a regular Cartesian coordinate system. Finally, in the [0º,36º] regime we examine Ψ, another well-accepted structural measure that has shown strong correlations with atoms' non-affine displacements in amorphous solids [28, 47]. As seen by the yellow triangles in **Fig. 2b**, there is a good linear correlation between Ψ and $\bar{E}_A$. To briefly summarize, among the 5 structural indicators considered in the present study, the $E_{IS}$ and $\Psi$ are correlated very well with the system's $\bar{E}_A$; while the other three measures, including $\rho$, $s_2$, and $|\vec{u}|$, only show weak correlations with $\bar{E}_A$.

It is worth noting that, the hereby considered $\bar{E}_A$ merely represents the simple arithmetic mean calculation of the $E_A$ distribution, However, in real physical and thermally activated processes, the Boltzmann weight factor, $\exp[-E_A/k_BT]$, has to be considered [18, 32], which renders the calculation of $<E_A>$ $T$-dependent. Therefore, it is more physically meaningful to obtain the probability distribution function $P(E_A)$, rather than the simple arithmetic mean value, $\bar{E}_A$. The reason is that, if two samples have the same $P(E_A)$ then they will always have the same performance; however, if they have the same $\bar{E}_A$ then they do not necessarily possess the same $P(E_A)$, which yield different behaviors under various conditions. To this end, in what follows we will investigate which structural indicator could potentially best describe $P(E_A)$ with the least uncertainty. As illustrated in **Fig. 3a**, to quantify the $E_A$ spectra difference between two samples here we define a $\delta$ function as $\delta_{i,j} = \int [P_i(E_A) - P_j(E_A)]^2 dE_A$, where $P_i(E_A)$ and $P_j(E_A)$ represent the spectra of samples #$i$ and #$j$, selected from the previously obtained dataset in **Fig. 2a**. In **Fig. 3b-f** we map the calculated $\delta_{i,j}$ into the 5 different structural spaces. By systemetrical analyzing the contour maps (see the Supporting Information), the distributions of $\delta_{i,j}$ in **Fig. 3e** and **Fig. 3f** suggest that $E_{IS}$ and $\Psi$ are still the two best candidates correlating with $P(E_A)$. We hypothesize that $E_{IS}$ might be a more viable option to express $P(E_A)$ for MG samples with similar density due to the relatively low computational costs. It is worth noting that, some recent attempts to use $E_{IS}$ as the structural parameter have successfully explained the MGs' exothermic features observed in experiments [32, 48], which provide further premises in support of our hypothesis.

Importantly, through the medium of $E_A$ spectra, we can expect a strong correlation between G and $E_{IS}$ or $\Psi$. As shown in Fig. 4a and b, the $\Psi$ presents much better correlation with G, which implies that the soft-mode parameter would probably be a more advantageous indicator for the activation energy spectra in metallic glass. It can be well explained by the intrinsic correlation between the soft-mode parameter and the Debye-Waller factor. On short timescales, the local Debye-Waller factor $\alpha_i$ is primarily determined by local structures that can be employed as a structural parameter as well. As proposed by Tong and Xu[47] based on the equipartition hypothesis, the soft-mode parameter for individual particles $\Psi_i = \sum_{j=1}^{dN-d} \frac{1}{\omega_j^2} |\vec{e}_{j,i}|^2$ is statistically proportional to the single-particle Debye-Waller factor $\alpha_i$ in glass systems. Previous simulations[38] have shown that the short-time local positional fluctuation $\alpha_i$ provides a good predictor of metallic glass properties at low temperature, such as shear modulus G. Thus, the soft-mode parameter $\Psi = \sum_{i=1}^{N} \Psi_i$, where N is the number of atoms, could be a good predictor of shear modulus G and the activation energy spectra in metallic glass. Moreover, $\Psi$ is mainly contributed by the atoms with larger $\Psi_i$ which are dominated by the low-frequency modes. It means that the parameter $\Psi$ quantitatively reflects the defective local structures in metallic glass. The strong correlation between G and $\Psi$ implies that the physical properties of metallic glass are sensitive to the geometrically unfavored Voronoi motifs rather than the favored Voronoi motifs, such as icoshedron[35-36].

In summary, we employ the ART method to directly probe the $E_A$ spectra for a number of $Cu_{50}Zr_{50}$ MG samples that are prepared at different pressure and cooling rate conditions over a broad range of parameter space. It is found that a sample's shear modulus is strongly correlated with its arithmetic mean of $E_A$ spectra rather than the local geometrical ordering. By quantifying the differences between each pair of obtained $E_A$ spectra and mapping them into various widely investigated representative parameter spaces, we demonstrate that the soft mode parameter Ψ has the greatest potential to determine a sample's $E_A$ spectrum. Through the medium of PEL, the soft mode parameter Ψ can be proposed as a good predictor of metallic glass properties, such as shear modulus G. Furthermore, the capability of establishing a quantitative description on $P(E_A)$ is critical, because as discussed above, it determines many macroscopic thermo-mechanical properties of amorphous solids [18, 31-32]. Although the present study does not yet give an explicit formulism of $P(E_A)$, it provides effective guidance towards such a goal. The hereby obtained correlations and uncertainty quantifications, in conjunction with machine learning tools, are expected to enable an analytical and accurate expression of $P(E_A)$ in the future.

**SIMULATION METHODS**

Molecular dynamics simulations have been performed by utilizing the open-source code-LAMMPS[49]. A 4000-atom simulation box, with periodic boundary conditions applied in all three directions, is first equilibrated at a supercooled liquid state with the temperature of 1500K and volume of 43×43×43Å$^3$. Then the glass samples are prepared by systematically tuning the hydrostatic stress (0, 5, 10, 15, 20 GPa) and the cooling rate ($10^{10}$, $10^{11}$, $10^{12}$, $10^{13}$ K/s) with the constant number, pressure, and temperature (NPT) ensemble. Then, the activation-relaxation technique (ART) method [50-51] was employed to probe the PEL of the prepared glassy samples and thereby its $E_A$ spectrum. To initiate the PEL sampling, small random perturbations with a total displacement of 0.5 Å are introduced to a cluster of atoms within the first nearest neighbor distance. Then, the ART searching algorithm will drive the system to leave the current basin (with the curvature criterion of -0.01eV/Å$^2$) and identify the connecting saddle states (with the force convergence criterion of 0.05 eV/Å). By statistically repeating such perturbation and search processes with various random seeds and over different positions in the system, the representative $E_A$ spectrum can thus be obtained (see Supporting Information for more details).

**Acknowledgments**

This work is supported by (S.Z. and P.G.) the NSF of China (Grants No. 51571011 and U1930402 ) and the MOST 973 Program (No. 2015CB856800). S.Z. and P.G. acknowledge the computational support from the Beijing Computational Science Research Center(CSRC). C.L. and Y.F. acknowledge the support from the U. S. Army Research Office under Grant No. W911NF-18-1-0119.

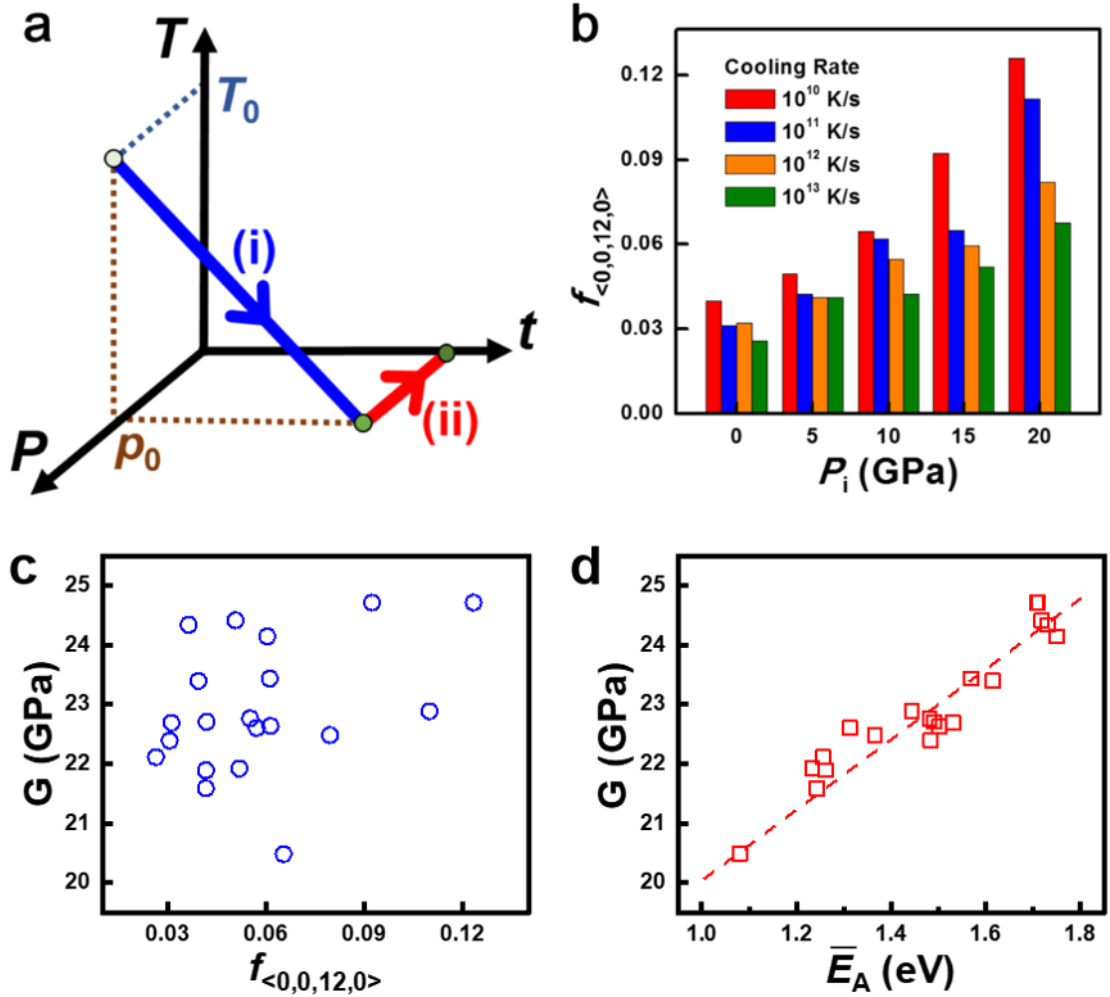

Figure 1 (a) Schematic diagram of the two-steps preparation process for each MG sample: (i) From $T_0$ quenched to 1K with different external pressures $P_i$ and cooling rates (ii) Released the pressures to 0 Gpa and kept the temperature at 1K and relaxed the samples for 10ns. (b) The fraction of icosahedron cluster in investigated MG samples. The correlation between the shear modulus G and the $f_{<0,0,12,0>}$ (c) and the the arithmetic mean of $E_A$ spectra, $\bar{E}_A$(d).

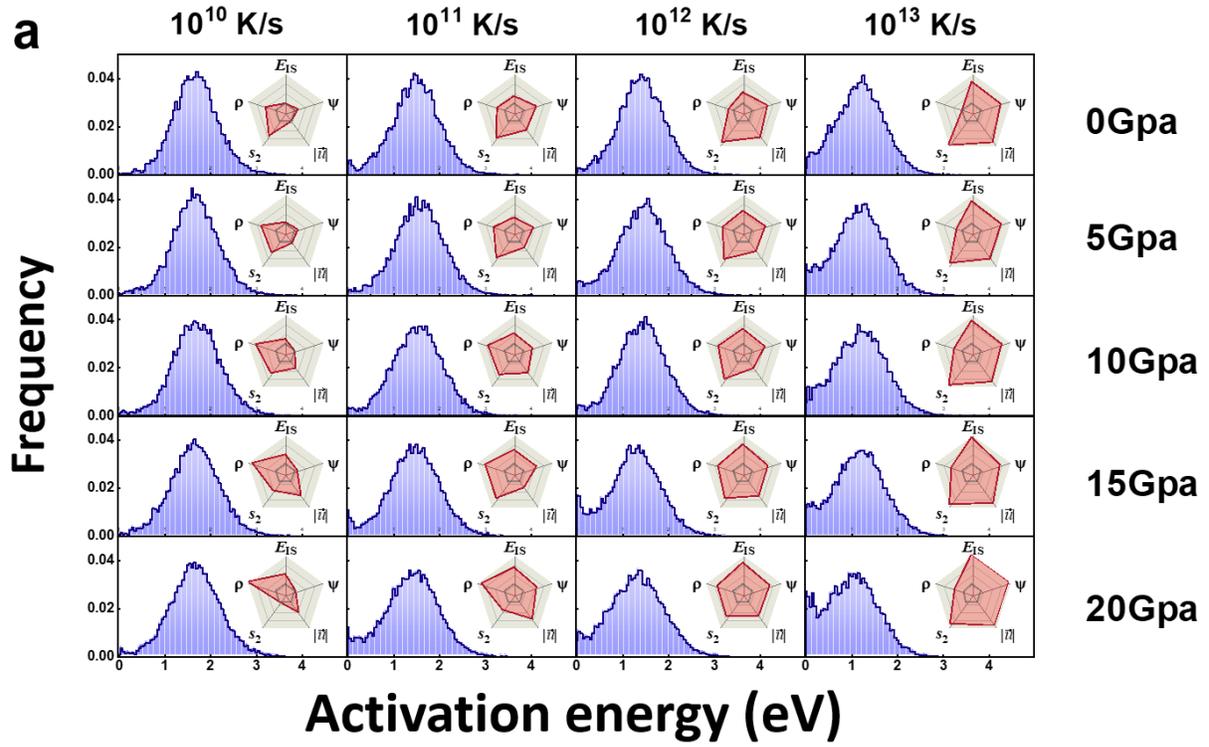

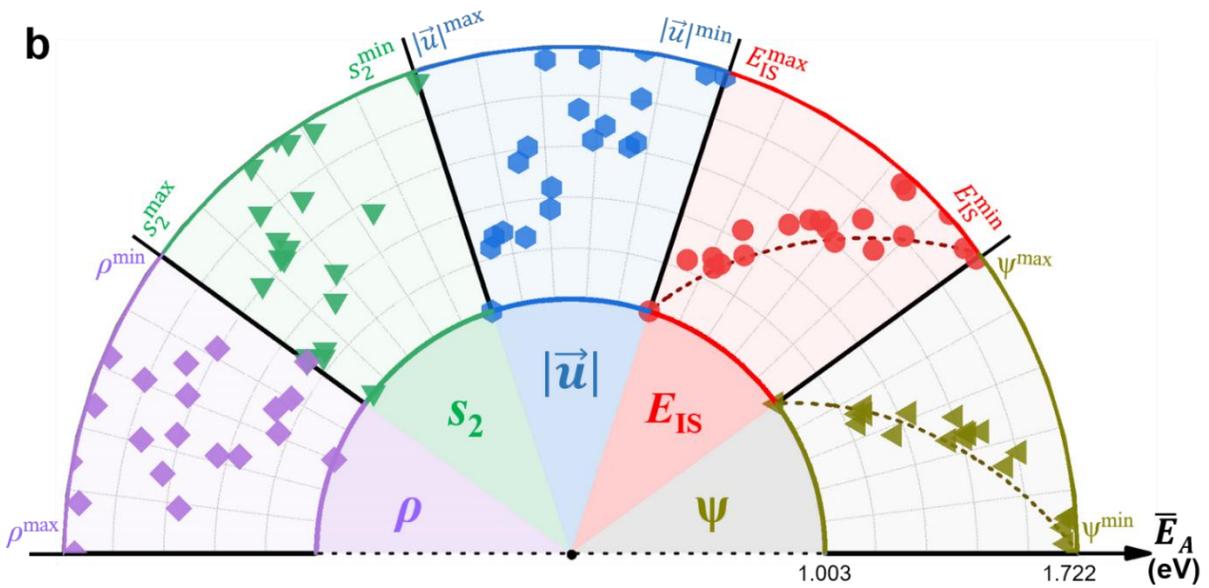

Figure 2 (a) The spectrum of $E_A$ for systems of various $P_i$ and cooling rates. Each panel's inset with the radar plot shows the sample's $E_{IS}$, $\Psi$, $|\vec{u}|$, $s_2$ and $\rho$, which scaled with the maximum value (raw data are shown in Fig.S2). (b) The average value of each parameter $vs$ $\bar{E}_A$ in polar coordinates.

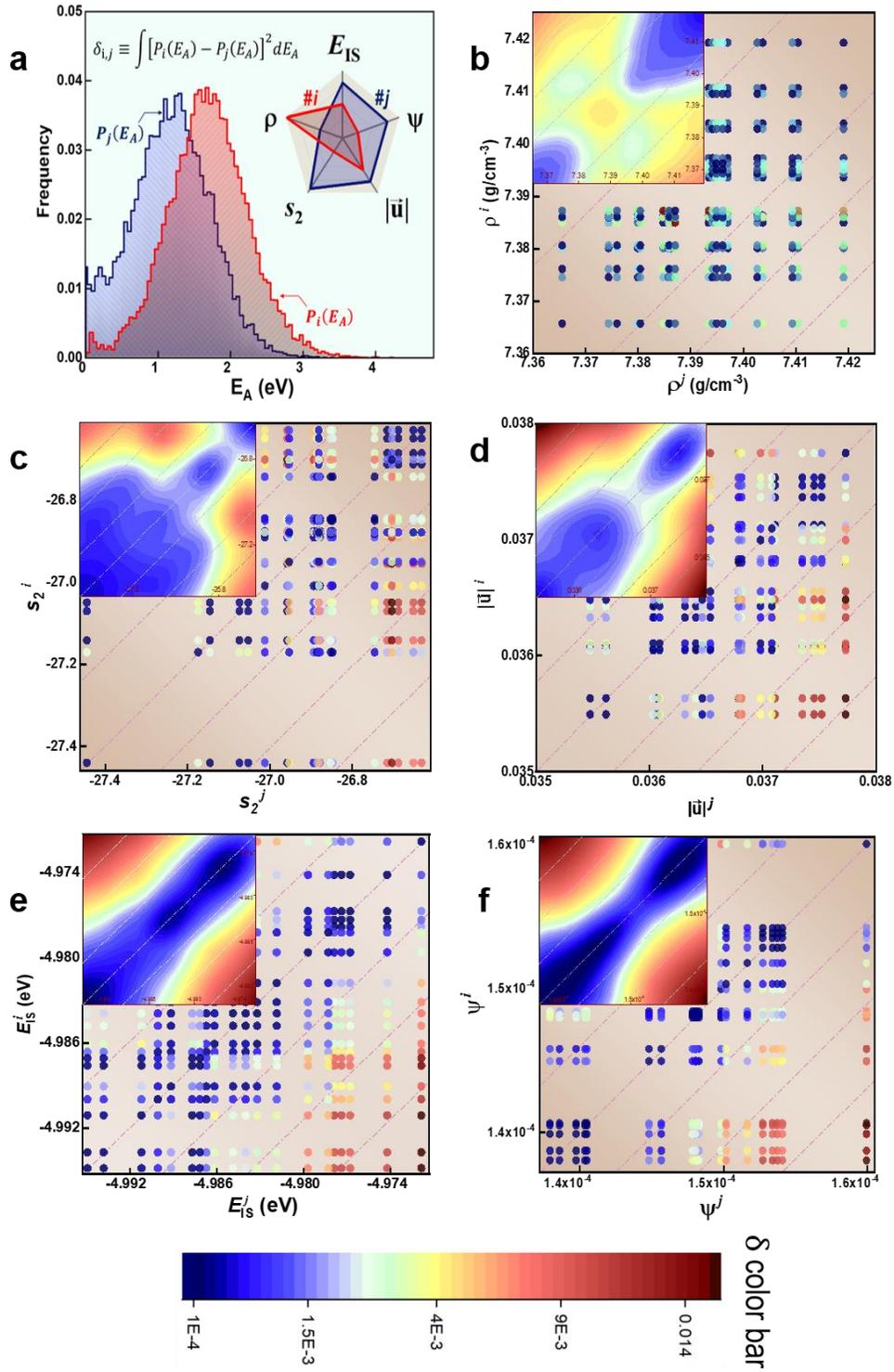1919

Figure 3 (a) The $E_A$ spectrum for the *i* and *j* systems. The difference of the $E_A$ spectra between *i* and *j* systems is defined as $\delta_{i,j}$. (b-f) Respective variations of $\delta_{i,j}$ dependence on $\rho$, $s_2$, $|\vec{u}|$, $E_{IS}$, and $\Psi$. The inset maps are contour maps of $\delta_{i,j}$.

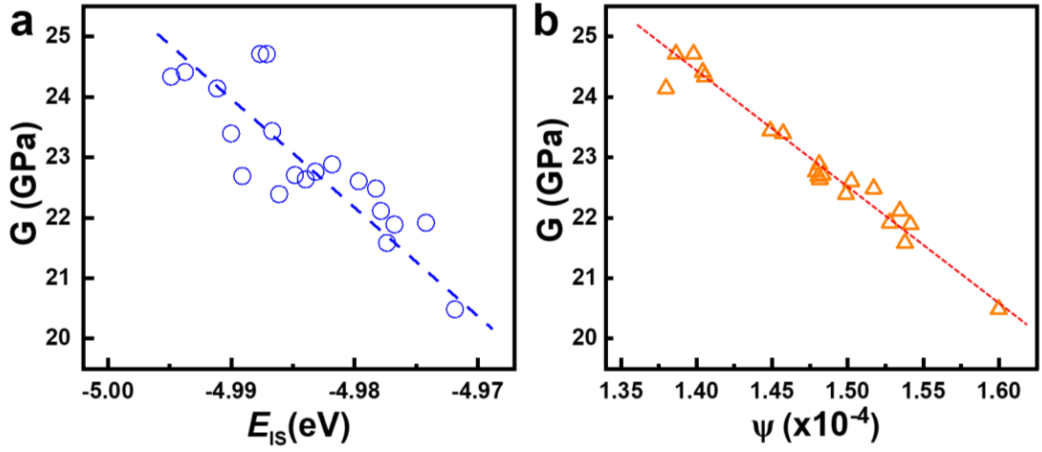

Figure 4 (a) The correlation between the shear modulus G and the $E_{IS}$ (a) and $\Psi$(d).

# Soft mode parameter as an indicator for the activation energy spectra in metallic glass

*Shan Zhang [a], Chaoyi Liu[b], Yong Yang[c], Yue Fan [b,], Pengfei Guan [a,*]*

[a] *Beijing Computational Science Research Center, Beijing, China*

[b] *Department of Mechanical Engineering, University of Michigan, Ann Arbor, USA*

[c] *Centre for Advanced Structural Materials, Department of Mechanical and Biomedical Engineering, City University of Hong Kong, Hong Kong, China*

## 1  $E_A$ spectrum and the arithmetic mean $\bar{E}_A$

**$E_A$ spectrum**. To obtain the **$E_A$ spectrum** of a given sample, we applied 10 activation-relaxation technique (ART) searches for each group of atoms. Thus, we can got the raw data by 40,000 ART searches for a given sample with 4000 atoms (center atom of a group). However, raw data contain failed and redundant searches. To acquire the accurate **$E_A$ spectrum, we employed two steps based on four parameters to screen the** failed and redundant searches.

Four parameters:

*the energy difference between the saddle and initial state, $E_{\text{sad-ini}} \equiv E_{\text{sad}} - E_{\text{ini}}$*
*the energy difference between the final and saddle state, $E_{\text{fin-sad}} \equiv E_{\text{fin}} - E_{\text{sad}}$*
*the energy difference between the final and initial state, $E_{\text{fin-ini}} \equiv E_{\text{fin}} - E_{\text{ini}}$*
*the distance between the final state and initial state, $\Delta \equiv \left[ \sum_{i=1}^{\#\text{atoms}} \left( \vec{r}_{\text{fin}}^{\,i} - \vec{r}_{\text{ini}}^{\,i} \right)^2 \right]^{1/2}$*

Two steps:

For each ART search,

1) 1) if $E_{\text{sad-ini}} < 0$ or $E_{\text{fin-sad}} > 0$, this event is a failed one and we will remove it. And then, if its $E_{\text{fin-ini}} < 0.02(\text{eV})$ and $\Delta < 1(\text{Å})$, we will remove it(because in this event, the final state is identical to the initial state). After these steps, we have removed the failed and the final state the same as the initial searches.

2) And then we go into the process of removing the redundant searches. For any pair of #*m* and #*n* in the search that after steps 1), if the $|\Delta^m - \Delta^n| < 0.1$(Å), and $|E^m_{sad-ini} - E^n_{sad-ini}| < 0.01$(eV), and $|E^m_{fin-ini} - E^n_{fin-ini}| < 0.005$(eV), we define them redundant and remove the redundant ones.

After these two removing steps, we have ~25,000 effective searches (in different samples, the number will be different ) with activation energy $E_A$. Finally, we can obtain the distribution of $E_A$, as the **E_A spectrum**, for each sample.

**The arithmetic mean $\bar{E}_A$.** The $\bar{E}_A$ is the arithmetic mean of all effective searches' activation energy $E_A$ in each sample.

$$\bar{E}_A = \frac{1}{N_{event}} \sum_{i=1}^{N_{event}} E_A(i)$$

where $N_{event}$ is the total number of valid events (~25000).

## 2  The calculated total radial distribution functions (RDFs)

For more information about the effects of different pressures and cooling rates on static structures, the total radial distribution functions RDFs of 20 samples prepared at four different cooling rates($10^{10}$ to $10^{13}$K/s) and five different pressures(0 to 20Gpa) are presented in Fig.S1 By comparing samples with varying cooling rates, we found that the changes in RDFs are not significant with the cooling rates. And the effects of stress are even more pronounced, although the external pressure has been released. At the same cooling rate, as the pressure increases, the shoulders of the first peak become gradually smooth and the splitting of the second peak becomes more pronounced. This shift shows that although the pressure is released, the pressure effect still retains some local geometric structure changes. The cooling rate has some impact on peak density. Combined with the above two factors, we can obtain 20 samples with very different local geometric structures.

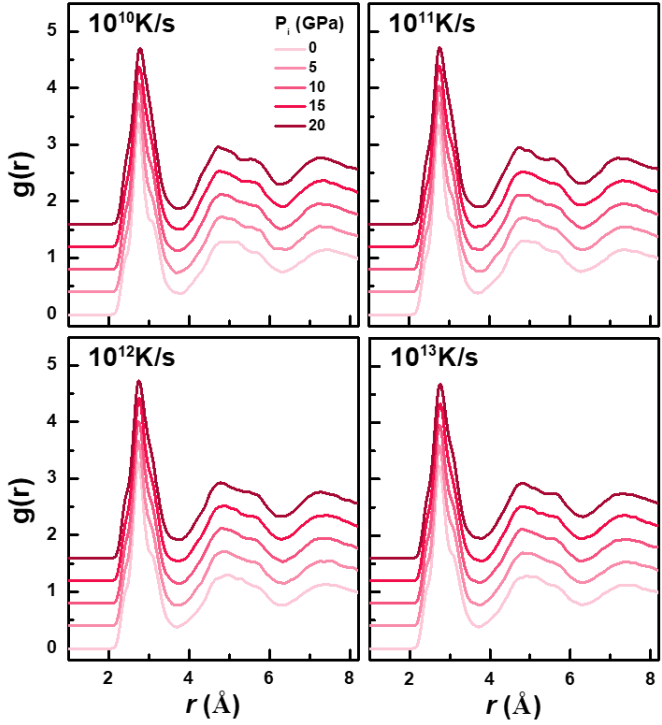

Fig.S1 The calculated radial distribution functions $g(r)$ of the samples at $10^{10}\sim10^{13}$K/s cooling rates with various pressures (0~20Gpa).

## 3  Local physical parameters at the atomic level

**Inherent structure energy $E_{IS}$:** Inherent structure energy $E_{IS}$ is defined as the potential energy of the local minimum configuration of the potential energy landscape[1]. When the samples were prepared, we use the conjugate gradient (CG) algorithm method to get the minimized configuration and the intrinsic structure energy (per atom).

**Local structural entropy $s_2$:** From the pair correlation function g(r), we use the defined local structural entropy $s_2$:

$$s_{2,i} = -1/2 \sum_v \rho_v \int d\vec{r} \{g_i^{\mu\nu}(\vec{r}) \ln g_i^{\mu\nu}(\vec{r}) - [g_i^{\mu\nu}(\vec{r}) - 1]\}$$

Where $\rho_\nu$ is the number density of ν particle, ν, μ is the particle types. Many works indicated the value of $s_{2,i}$ is correlated well with the slow dynamics [2] and the heterogeneous dynamics[3-4] of MGs. The average value $s_2$ of the sample:

$$s_2 = \langle \sum_{i=1}^{N} s_{2,i} \rangle$$

where N is the number of atoms in the system.

**Soft mode parameter Ψ :** The structure order parameter Ψ is the participation ratio of low-frequency mode, correlating with the low-temperature dynamics and local surroundings Using the classical approximation and equipartition of energy to all vibration modes, the mean square vibrational amplitude of particle *i* is defined by

$$\Psi_i = \sum_{j=1}^{dN-d} \frac{1}{\omega_j^2} |\vec{e}_{j,i}|^2$$

Where $\vec{e}_{j,i}$ is the polarization vector of particle *i* of mode *j* by the vibrational frequency ω. Except for the *d* zero-frequency modes, these are *dN – d* modes in the system with N atoms. The low-frequency modes dominate the value of $\Psi_i$. It has a high correlation with Debye-waller factor and reflects the fast local dynamic properties of MGs[5]. Particles with large $\Psi_i$ have more opportunities for vibration and rearrangement. Given the overall, we used the average Ψ:

$$\Psi = \langle \sum_{i=1}^{N} \Psi_i \rangle$$

where *N* is the number of atoms in the system.

**Local configurational anisotropy $|\vec{u}_i|$ :** The local configuration anisotropy is defined as the displacement from particle center to the corresponding Voronoi cell centroid. We usually use the Voronoi cell structures to define the local neighborhood of atoms in MGs. In this work, we set the modulus of the vector $|\vec{u}_i|$ [6] is the displacement from the particle *i*'s center to the average position of its Voronoi cell vertices. And it corresponds to the rearranging region space trend[6-8].

$$|\vec{u}| = \langle \sum_{i=1}^{N} |\vec{u_i}| \rangle$$

where the sample has N atoms.

## 4 The raw data of Figure 2b.

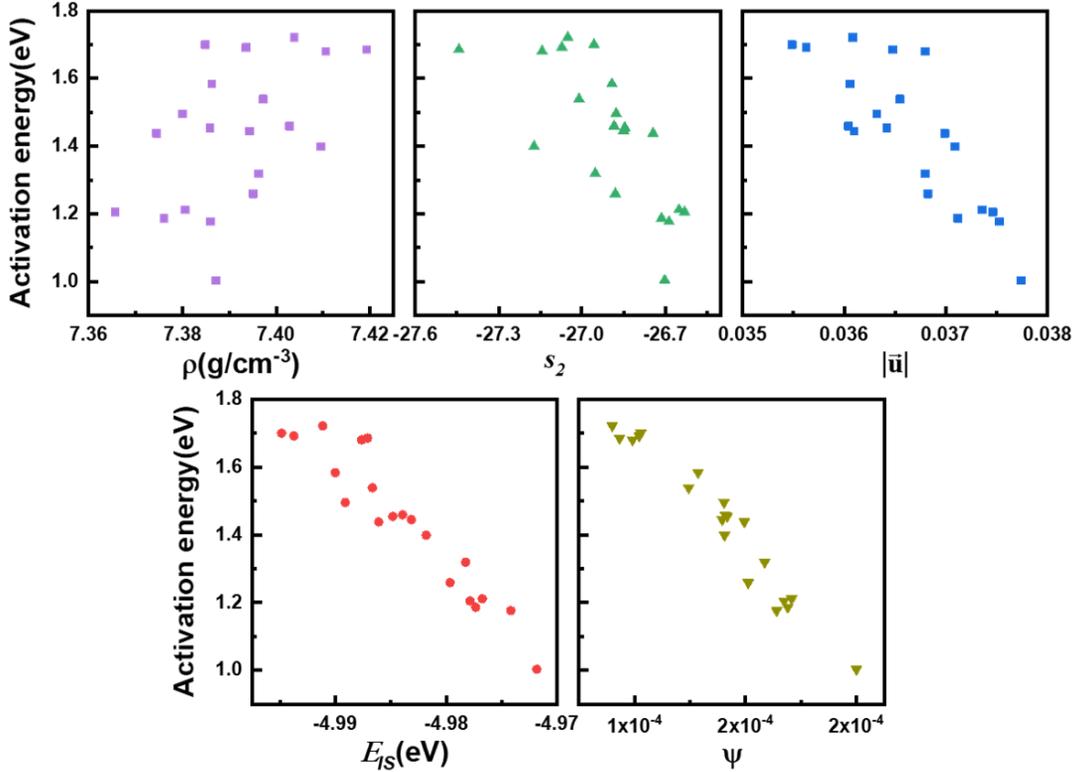

Fig.S2 The samples' average activation energy $\bar{E}_A$ dependence on $\rho$, $E_{IS}$, $s_2$, $\Psi$, and $|\vec{u}|$

## 5 Details about the data analysis of Fig. 3.

As shown in Fig.3, in each panel there are 400 (=20*20) data entries, which are colored according to its δ value marked by the color bars. Note that not all the 400 points are independent, because the symmetry (*i.e.* $P_{ij} = P_{ji}$) and zero diagonals (*i.e.* $P_{ii} = 0$) have to be satisfied due to the definition of Eq. (1). For the convenience of analyses, in the inset plots those discrete data points are converted into quasi-continuum contour maps using the thin-plate spline (TPS) interpolation algorithm[9]. If there does exist one structural indicator that is able to characterize the $P(E_A)$ with a minimum blurriness, then the contour map should possess the following two features: At first, the largest δ values should be concentrated near the up-left and bottom-right corners, because those extreme off-diagonal regions are where the structural differences between samples #$i$ and #$j$ are maximized; Secondly, the transition from the blue regime to the red regime should be continuous and smooth, so that it would ease the future development of an analytical expression of $P(E_A)$. Following such spirit, **Fig. 3e** and **Fig. 3f** stand out among their peers, suggesting that $E_{IS}$ and Ψ are still the two best candidates correlating with $P(E_A)$.